\documentclass[structabstract]{aa}  
\usepackage{graphicx}
\usepackage{natbib}
\begin{document}
  \title{On the ultraviolet signatures of small scale heating in coronal loops}

  \subtitle{}

  \author{S. Parenti
         \inst{1}
         \and
         P. R. Young \inst{2}\thanks{Present address: Code 7673,
           Naval Research Laboratory, Washington, DC 20375, U.S.A.}
         }

  \offprints{S. Parenti}

  \institute{Royal Observatory of Belgium, 3 Av. Circulaire, 1180 Bruxelles, Belgium \\
             \email{s.parenti@oma.be}
        \and
          Rutherford Appleton Laboratory, Chilton, Didcot, Oxfordshire, OX11 0QX, U.K.            
            }

  \date{}

% \abstract{}{}{}{}{} 
% 5 {} token are mandatory
\abstract{}
 % context heading (optional)
%\abstract   {}
%  {}   aims heading (mandatory)
  {Studying the statistical properties of  solar ultraviolet emission
  lines  could provide information about the nature of  small scale coronal
  heating. We expand on previous work to investigate
  these properties. We study whether the predicted
  statistical distribution of ion emission line intensities produced by a
  specified heating function is affected by the isoelectronic
  sequence to which the ion belongs, as well as the characteristic temperature at which it was
  formed (as found previously). Particular emphasis is placed on
  the strong resonance  lines belonging to the lithium isoelectronic
  sequence. Predictions for emission lines observed by existing
  space-based UV spectrometers are given.  The effects on the statistics of a line when
   observed with a wide-band imaging instrument rather than a
  spectrometer are also investigated.}
%   methods heading (mandatory)
  {We use a hydrodynamic model to simulate the UV emission of a loop
  system heated by nanoflares on small, spatially unresolved
  scales. We select lines emitted at similar temperatures but
  belonging to different isoelectronic groups:  \ion{Fe}{ix} and
  \ion{Ne}{viii}, \ion{Fe}{xii} and \ion{Mg}{x}, \ion{Fe}{xviii}, \ion{Fe}{xix} and  \ion{Fe}{xxiv}.}
%   results heading (mandatory)
  {Our simulations confirm previous results that almost all
  lines have an intensity distribution that follows a power-law, in a
  similar way to the heating function. However, only the high temperature
  lines best preserve the heating function's power law index (\ion{Fe}{xix} being the best ion in the case presented here). The Li
  isoelectronic lines have different
  statistical properties with respect to the lines from other
  sequences, due to the extended high temperature tail of their
  contribution functions. However, this is not the case for
  \ion{Fe}{xxiv} which may be used as a diagnostic of the coronal
  heating function.  We also show that the power-law index of the
    heating function is effectively preserved when a line is observed
    by a  wide-band imaging instrument rather than a spectromenter.}
%   conclusions heading (optional), leave it empty if necessary 
%  {The results of this work give a warning when a statistical study is done on the EUV line with the purpose of deriving information on the heating. }
{}

  \keywords{Sun: UV radiation -- Sun: corona -- Plasmas -- Methods: statistical
              }
%\authorrunning{}

\titlerunning{Small scale heating in coronal loops}

  \maketitle
%
%________________________________________________________________

\section{Introduction}

VUV and X-ray small-scale brightenings are often detected in
 images or spectra of the solar atmosphere
 \citep[e.g.][]{berghmans98, aletti00,
   aschwanden02, christe08}. These observations and theoretical
 considerations indicate that these brightenings may be one
 manifestation of small-scale impulsive heating acting on the
 unresolved fine scale of the corona. Modelling unresolved small-scale brightenings can help, for example, to explain the long
 lifetimes of active region loops. Understanding the properties of this heating and its contribution to global coronal heating are among the most challenging questions in solar physics.  important starting point is identifying key 
observables that can diagnose proposed heating functions.

The study of the frequency distribution of the intensity of coronal VUV
emission lines is one method for investigating this
problem. The distributions derived from observations with a variety of
instruments and data sets show power laws
 with indices mainly between $-1.5$ and $-2$
\citep[e.g][]{aschwanden05}. Converting the frequency distribution into
a distribution of thermal energy for the heating events yields a similar power-law, and this is assumed to represent the power-law index of the
(unknown) coronal heating function.
% The conversion to thermal energy shows a similar property.  
% These statistical properties are then used to infer the same property to the (unknown) coronal heating function. 
However, behind this method there is the assumption that the
energy conversion mechanisms do not modify the original heating energy distribution, so that the measured events distribution is equivalent to the heating distribution.

\cite{parenti06} investigated this aspect using a forward
 modelling approach: a statistical model of
coronal heating was posited and properties of the predicted emission
line intensities and thermal energies were deduced. 
Their main result indicated that only high
temperature lines, formed when the main cooling mechanism in the loop was conduction, are reliable. 

These authors showed that the statistical properties of the heating function are conserved by the line intensity statistical distribution,
only during such a period. The lines analyzed by these authors did not
belong to the  lithium isoelectronic sequence.

As with many other
ions formed in the solar atmosphere, the Li isoelectronic ions are formed  principally
over a narrow temperature interval ($\approx 0.3$~dex in
$\log\,T$), although, unlike most other ions, they also have a tail in
their temperature distributions that extends to high temperatures
(e.g. Fig.~\ref{GT}), which leads to the ions sampling a wider
temperature range than most other ions. 

In the present work, the results of Parenti et al. (2006) will be extended to investigate the following questions:
\begin{itemize}
\item How are the statistical distributions affected if the emission lines belong to the lithium-like isoelectronic sequence?

\item How will a statistical distribution measured from a wideband imaging instrument compare with a distribution from a single spectral line observed by a spectrometer?

\end{itemize}

The work presented here is particularly useful in the light of upcoming
 multi-channel,  high resolution solar VUV instruments. The
 Atmospheric Imager Assembly \citep[AIA,][]{golub06} is scheduled to be flown on the
 Solar Dynamics Observatory (SDO) in 2009 and has eight distinct
 channels, seven of which are narrow bandpass channels centred on strong
 emission lines. Solar Orbiter is planned to be launched in 2015 and
 VUV imaging and spectroscopic instruments -- the Extreme Ultraviolet
 Imager (EUI) and Extreme Ultraviolet Spectometer (EUS), respectively
 -- are part of the strawman
 payload \citep{hochedez07, young07a}.
% The work presented here is particularly useful in the light of the
% multi-channel high resolution FUV instruments that will soon fly (such
% as the SDO/AIA imager) or which are in preparation \cite[e.g. the
% spectrometer EUS, the imager EUI  on Solar Orbiter,][]{young07,
%   hochedez07}.
Due to the multiple hot channels present on SDO/AIA, it appears to be a
promising instrument, in terms of statistics, for observing small-scale heating events. 
In a more general context,
the plasma conditions in the solar atmosphere can be diagnosed if a set of lines formed at each layer of the solar atmosphere, can be detected.
 The coronal Li-like ions, as well as some of the \ion{Fe}{} ions studied here, have lines at long UV wavelengths that can be simultaneously observed with chromospheric and transition region lines (e.g. \ion{Fe}{xviii} and \ion{Fe}{xix} in the SUMER waveband). This is an important consideration for future spectrometers where we want access to all layers of the solar atmosphere with just two or three spectral bands \citep{young07a}. 

Due to the large number of current and planned VUV instruments, the present work concentrates on VUV lines and we do not investigate
the consequences of the small-scale brightening model on X-ray
emission. However, to our knowledge, soft X-ray microflare 
studies of this kind have not been completed previously, although  results from the Hinode Soft X-ray Telescope (SXT) may be forthcoming.

% , possibly because it is expected
% a very faint signal that is only accessible to high sensitive
% instruments. This could be a challenge for Hinode/XRT and the future
% soft X-ray instruments. 

% It is not the aim of this paper to investigate X-ray small scale emission. However, to our knowledge, soft-X ray microflare study of this kind have not been done, possibly because it is expected a very faint signal that is only accessible to high sensitive instruments. This could be a challenge for Hinode/XRT and the future soft X-ray instruments.

The paper is organised as follows.
Section 2 introduces the hydrodynamical model used in the simulations.
Section 3 describes how the synthetic spectra are built.
Section 4 provides the results, and conclusions are drawn in Sect. 5.

%__________________________________________________________________

\section{The model}

The hydrodynamical model used  was described in detail by
\cite{parenti06}, and we provide only a brief description.
%% For the following analysis we used the same hydrodynamic model used by \cite{parenti06}.
%% We give here a brief description. More details can be found in \cite{parenti06} and bibliography therein.
A coronal loop was modelled by a bundle of unresolved, identical
threads following the approach of \cite{cargill94}. Two thousand threads were
used each with a half length, $L$, of $10^9 ~\rm cm$, and a
cross-sectional area, $A$, of $2\times 10^{14}~ \rm cm^2$. At each
time $t$, a thread was described by a single temperature, $T$, and single density, $N$. The model simulates the heating-cooling cycle of each thread independently.  After the thread was heated impulsively, the cooling proceeded first by thermal conduction (at high temperatures) and then by radiation (at low temperatures). During the period dominated by conduction, the strand was filled with plasma by evaporation from the chromosphere, while it drains during the radiation phase.
The time evolution of the density and temperature in each thread was used to calculate the time-dependent synthetic spectra in the entire loop. The assumed energy loss was $2.5 \times 10^{-4}~ \mathrm{erg~ cm^{-3}~ s^{-1}}$.

\begin{figure}
  \centering
  \includegraphics[scale=0.5]{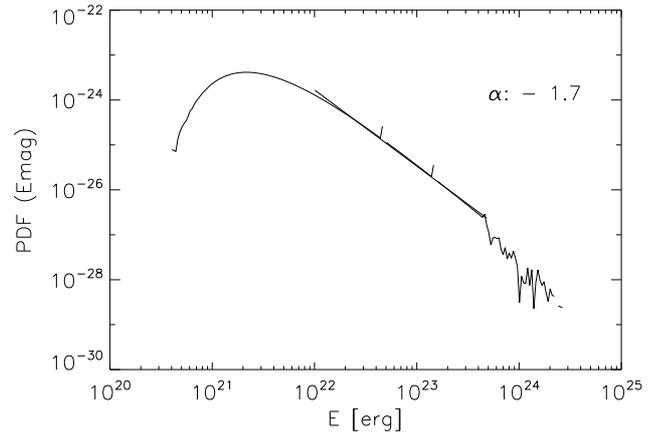}
  \caption{Probability Distribution Function of the heating function used for our simulations.}
             \label{heating_f}%
   \end{figure}

The heating function employed by \cite{parenti06} was derived from a
model of coronal turbulence developed by \cite{buchlin03}. For the present work,
a synthetic heating function was used,
%% We built a synthetic heating function (having comparable statistical
%% properties of the function used in \cite{parenti06})  
which simulated  a sequence of small energy-impulsive events ($\approx 10^{21} - 10^{24}~ \mathrm{erg}$), each involving only one thread at a time. 
The  heating function had a log-normal distribution with index $\alpha
= -1.7$,  over the energy range $\approx 10^{22} - 10^{24}~
\mathrm{erg}$ (Fig. \ref{heating_f}). 
Our use of this synthetic heating function is justifiable because power-law
distributions of observed energy events have been reported, and
theoretical models have also been shown to generate a power-law
distribution of energies \citep{parenti06}. A synthetic function can
also be easily modified, to allow a full investigation of model parameters.

%% Our aim is to simulate
%% nanoflares at the limits of current instrument capabilities that may have a turbulent origin. 
%% Heating function distributions deduced either by observations or by modeling  having a turbulent origin (the latter is, for example, the case for the work in \cite{parenti06}), have been found to follow a power-law distribution. Our function can well reproduce such a statistical property. Moreover, differently from what was used in our previuos work, it can also be used for a fully investigation of its parameters. This will be a target of our future investigations.

% For this reason we reduced the statistics in the highest energy part of the distribution. In fact we noticed that these high energy flares may play an important role in the statistics of the line intensities. 

%
%To make a more realistic energy distribution we imposed a drop in the power-law for energies greater than $4 \times 10^{24} mathrm{ergs}$.
%We then decided to cut the high energy part of the distribution by imposing an higher index to the power-low.

\section{The synthetic spectra}

To study the behaviour of the optically thin FUV-UV line intensities, it is helpful to consider the column Differential Emission Measure (DEM) distribution, which represents the amount of material at a given temperature along the line of sight ${h}$:

\begin{equation}
\mathrm{DEM}(T) = N^2 \left({d T}\over{d h}\right )^{-1} 
\end{equation}

\noindent where $N$ is electron density along $h$ at a given temperature ($T$).
This quantity is linked to a line intensity ($I$) for an optically thin line by: 
\begin{equation}\label{eq_dem}
I = \int_0^\infty \mathrm{ A(X)}~ G(T) ~  \mathrm{DEM}(T) ~  dT  ~~~ \mathrm{[erg ~  cm^{-2}~   s^{-1}~  sr^{-1}]}
\end{equation}
\noindent where 
$G(T)$ is the contribution function that contains the atomic physics
information for the emission line and is predominantly a function of temperature for most allowed transitions, and $\mathrm{A(X)}$ is the abundance of the element with respect to hydrogen. 

In our model, we assumed that the line of sight was perpendicular to the loop axes, so that the line of sight was given by the diameter of the strand multiplied by the number of strands emitting at the given temperature.
Under these conditions, the shape of the DEM can be represented by the distribution of $N^2$ versus $T$. (For details on the dependence of $\mathrm{DEM}(T)$ on the model parameters see \cite{parenti06} and references therein.)

\begin{figure}[h]
\includegraphics[scale=0.5]{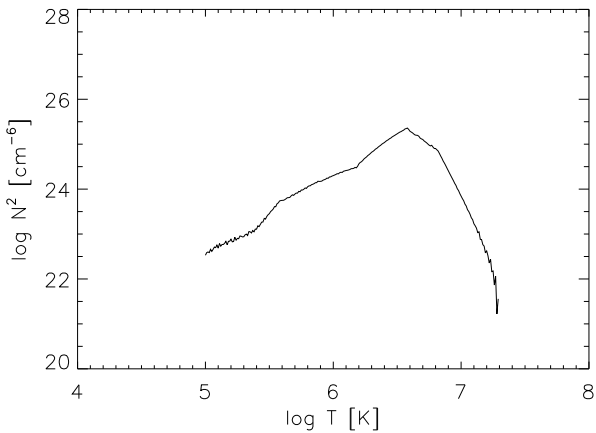}
\caption{ Variation in ${\log~ N^2}$ with ${\log~T}$ inside  the loop.}\label{figem}
\end{figure}

Figure \ref{figem}  shows the logarithm of ${N^2}$ as a function of the logarithm of temperature, integrated over the entire loop system (2000 strands) and simulation time ($10^5~ \mathrm{s}$). 
This plot helps to interpret our results. 

As mentioned earlier, in this model the cooling of a
strand proceeded in two separate phases: the conduction phase, where the strand is still at a high temperature; and the
radiation phase, for lower temperatures. For this reason, we are able to follow the cooling process in Fig. \ref{figem}.
The plasma evaporation from the chromosphere acting during the period dominated by  thermal conduction fills the high temperature part (on the right side from the peak) of the  ${N^2}$--${T}$ distribution. The lower temperature part of the distribution is filled during the radiation phase. The peak corresponds to the temperature location at which part of the strands are cooling by conduction and part by radiation.

One of the main results of 
\cite{parenti06} was that the intensity distribution of a spectral line
is representative of the heating function distribution, if the
formation temperature of the line is within the high temperature region
of the DEM distribution. This is due to the proximity with the
moment of heat injection, which produces radiative
emission that inherits the properties of the heating function.

For the case presented here, the DEM peaks at about $\log~T =
6.6$. This implies that lines emitted at lower temperatures
are formed while the cooling in the strand is dominated by
radiation. For $\log~T > 6.9$, the cooling is dominated by
conduction. An intermediate condition is found between these two
temperatures, in which the line emission loses all information about the
heating function.
%% Often the lines formed in this region of temperature completely lose information on the heating function.

Table \ref{tab1} lists the lines modelled in the present work, and their formation temperatures. These lines are observable with the instruments SOHO/SUMER, EIT and Hinode/EIS. Where possible, we chose two lines with similar formation temperatures but belonging to ions of different isoelectronic sequences. Lines belonging to the
Li isoelectronic sequence are highlighted. The SUMER
instrument 
on board SOHO is an
ultraviolet spectrometer observing in the wavelength range
500--1600~\AA\, which was described by \cite{wilhelm95}.  This is one of
the most useful bands in the VUV because it contains lines emitted from the chromosphere to the corona. 
Another SOHO instrument is EIT, which acquires solar
images in four wavelength bands by using multilayer optical coatings
\citep{delaboudiniere95}.
EIS is an
ultraviolet spectrometer on board the Hinode satellite operating in
the wavelength ranges 170--211~\AA\ and 246--292~\AA\ and was described
in \cite{culhane07}. 
%% The lines are currently observed by several instruments, such as SOHO/SUMER (\ion{Ne}{viii}, \ion{Mg}{x} at second order, \ion{Fe}{xiii}, \ion{Fe}{xix}), Hinode/EIS (\ion{Fe}{ix}, \ion{Fe}{xii}, \ion{Fe}{xxiv}).
%% As we will discuss in the next section, we also copared the results from the EIT 171 wave band, dominated by the \ion{Fe}{ix}, and the same line observed on Hinode/EIS. This in order to test the effect of a large band instrument on the PDF shape of the measured intensity.

\begin{figure}[th]
\includegraphics[scale=0.5]{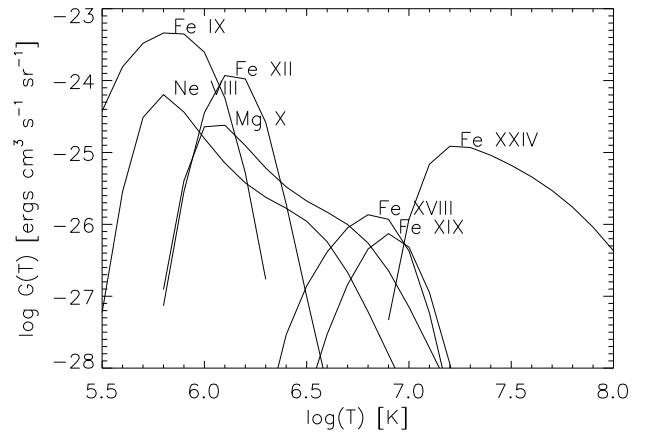}\\
\includegraphics[scale=0.5]{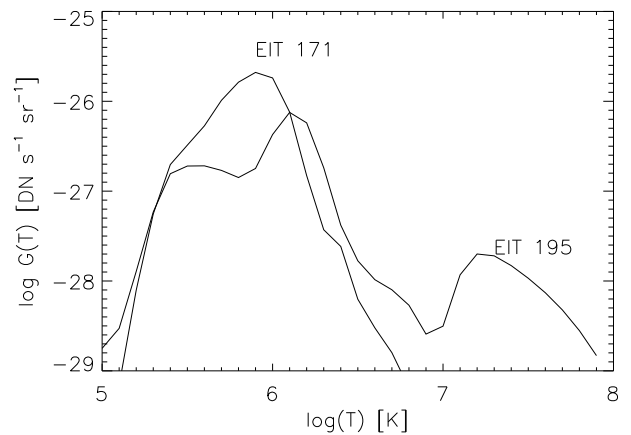}
\caption{Contribution functions for the spectroscopic lines (top) and the  temperature response function for the EIT 171 and 195 filters (bottom). 
}\label{GT}
\end{figure}

 \begin{table}[h]
     \caption[]{List of lines synthesised in this work, and with the logarithm of the temperature of their maximum emission. These lines can be observed by SOHO/SUMER (\ion{Ne}{viii}, \ion{Mg}{x}, \ion{Fe}{xviii},  \ion{Fe}{xix}), SOHO/EIT (\ion{Fe}{ix}, {\bf \ion{Fe}{xii}}) and Hinode/EIS (\ion{Fe}{ix}, \ion{Fe}{xii}, \ion{Fe}{xxiv})}
        \label{}
       \centering                          % used for centering table
        \begin{tabular}{c l l }        % centered columns (4 columns)
        \hline\hline                 % inserts double horizontal lines 
             $\log~(T_{\rm max})$ & Li-like &  Others \\
           \hline
          5.8  & \ion{Ne}{viii} 770.41 \AA& \ion{Fe}{ix} 171.07 \AA\\
          6.1 &  \ion{Mg}{x} 624.94 \AA & \ion{Fe}{xii} 195.12 \AA\\
           6.8 & & \ion{Fe}{xviii} 974.86 \AA \\
            6.9 &                     & \ion{Fe}{xix} 1118.06 \AA \\ 
          7.24 & \ion{Fe}{xxiv} 192.03 \AA &  \\
           \hline
   \end{tabular}\label{tab1}
  \end{table}
 \begin{figure*}[fth]
  \centering
 \includegraphics[scale=.5]{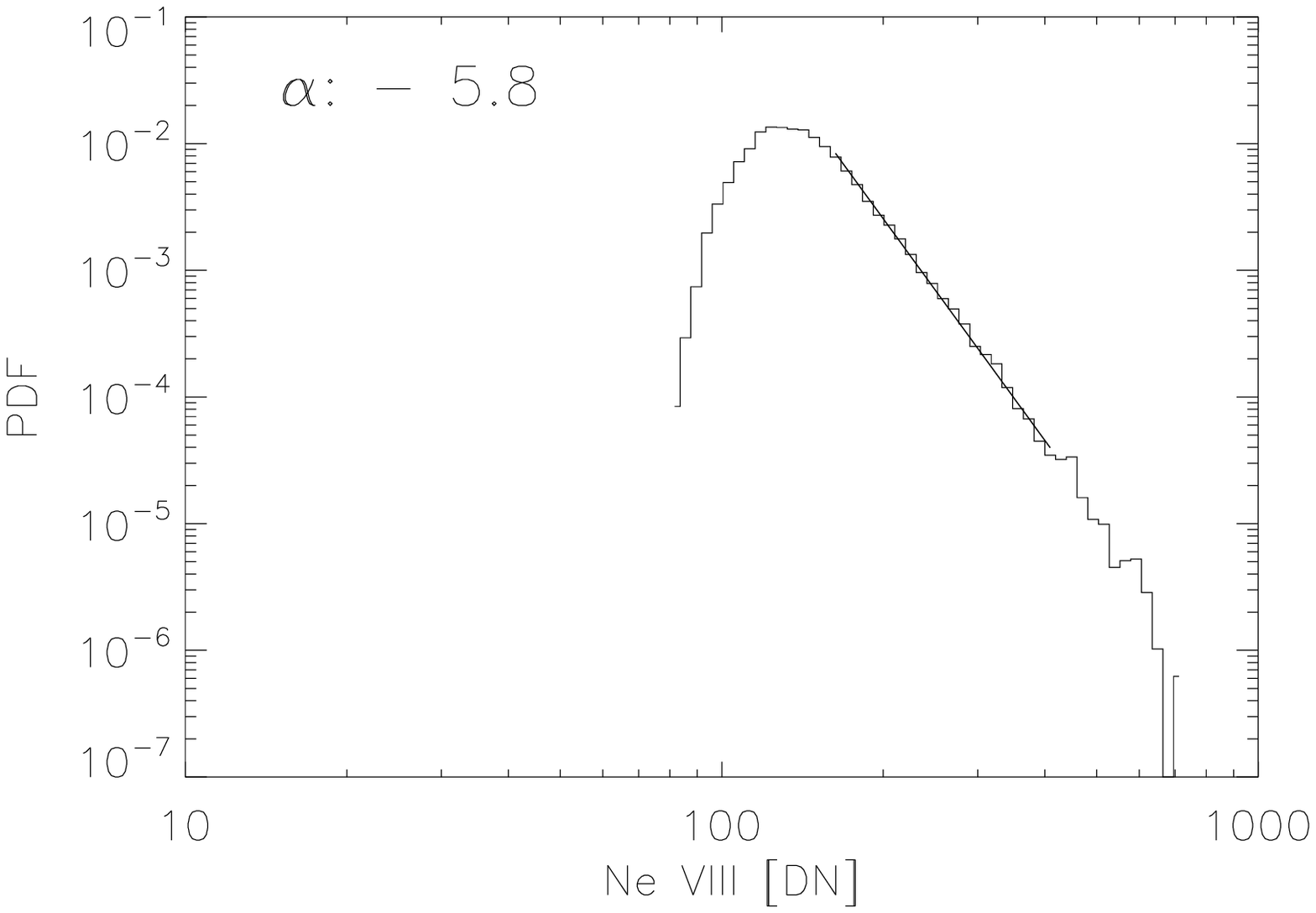}
 \includegraphics[scale=.5]{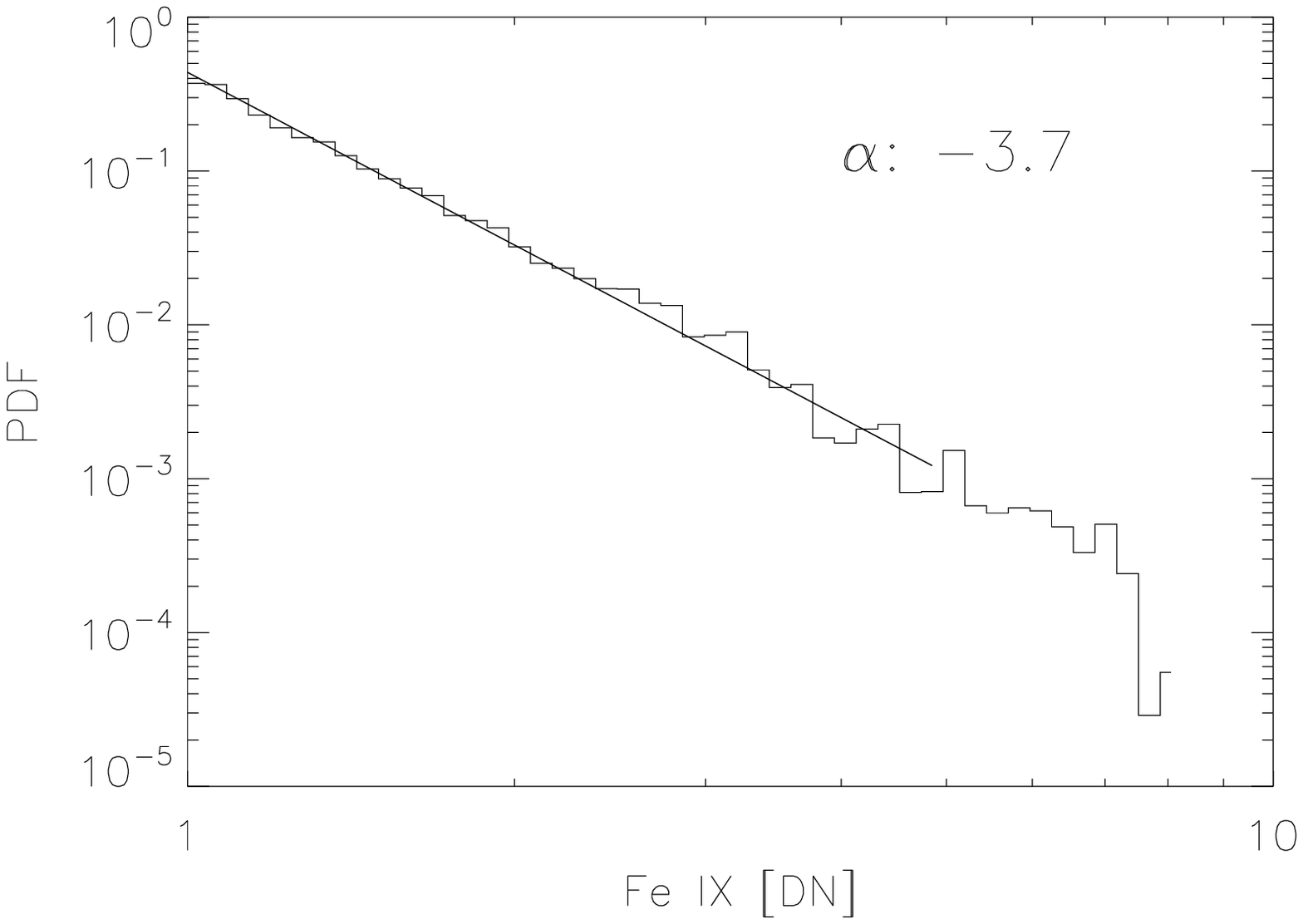}\\
 \includegraphics[scale=.5 ]{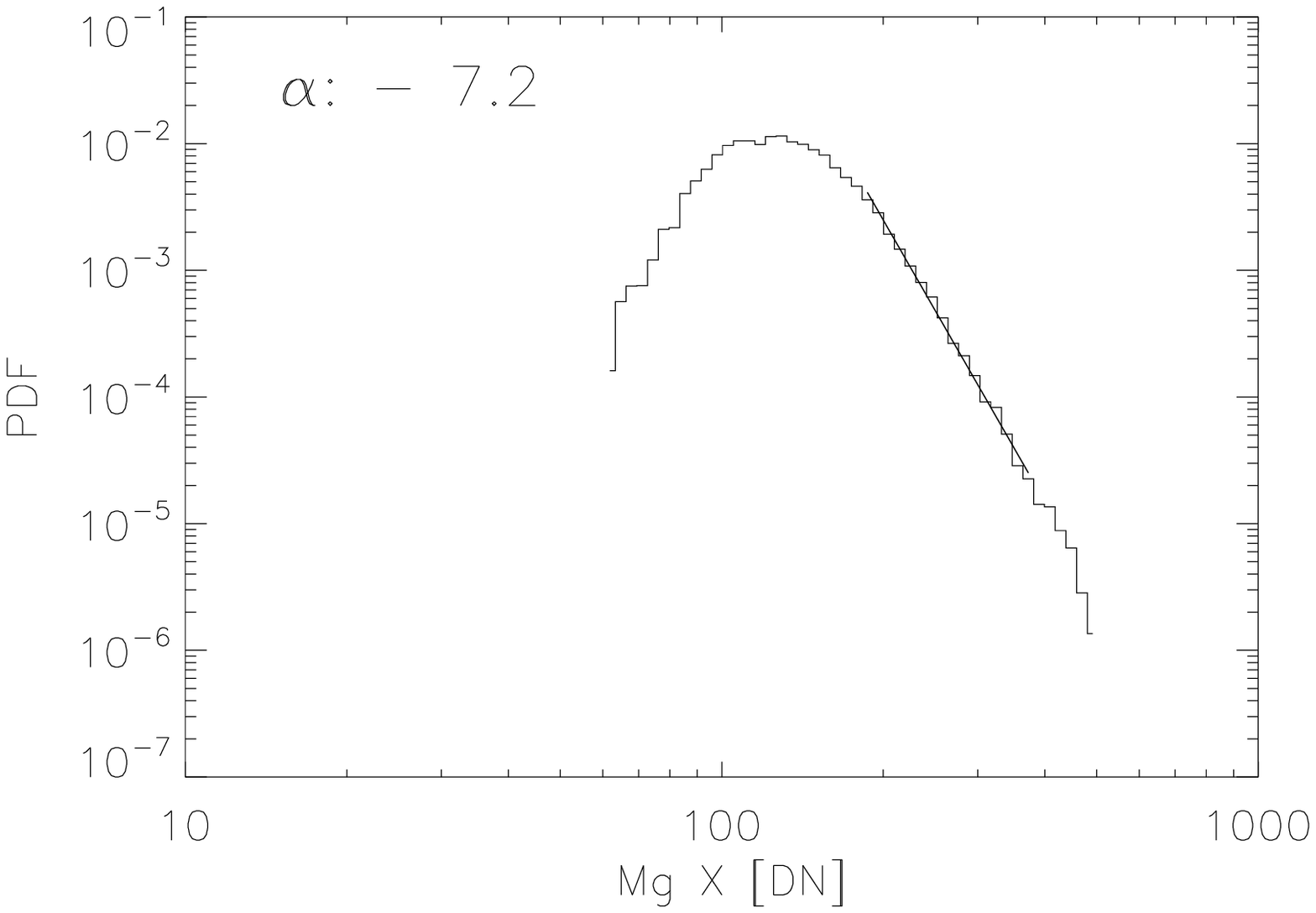}
 \includegraphics[scale=.5 ]{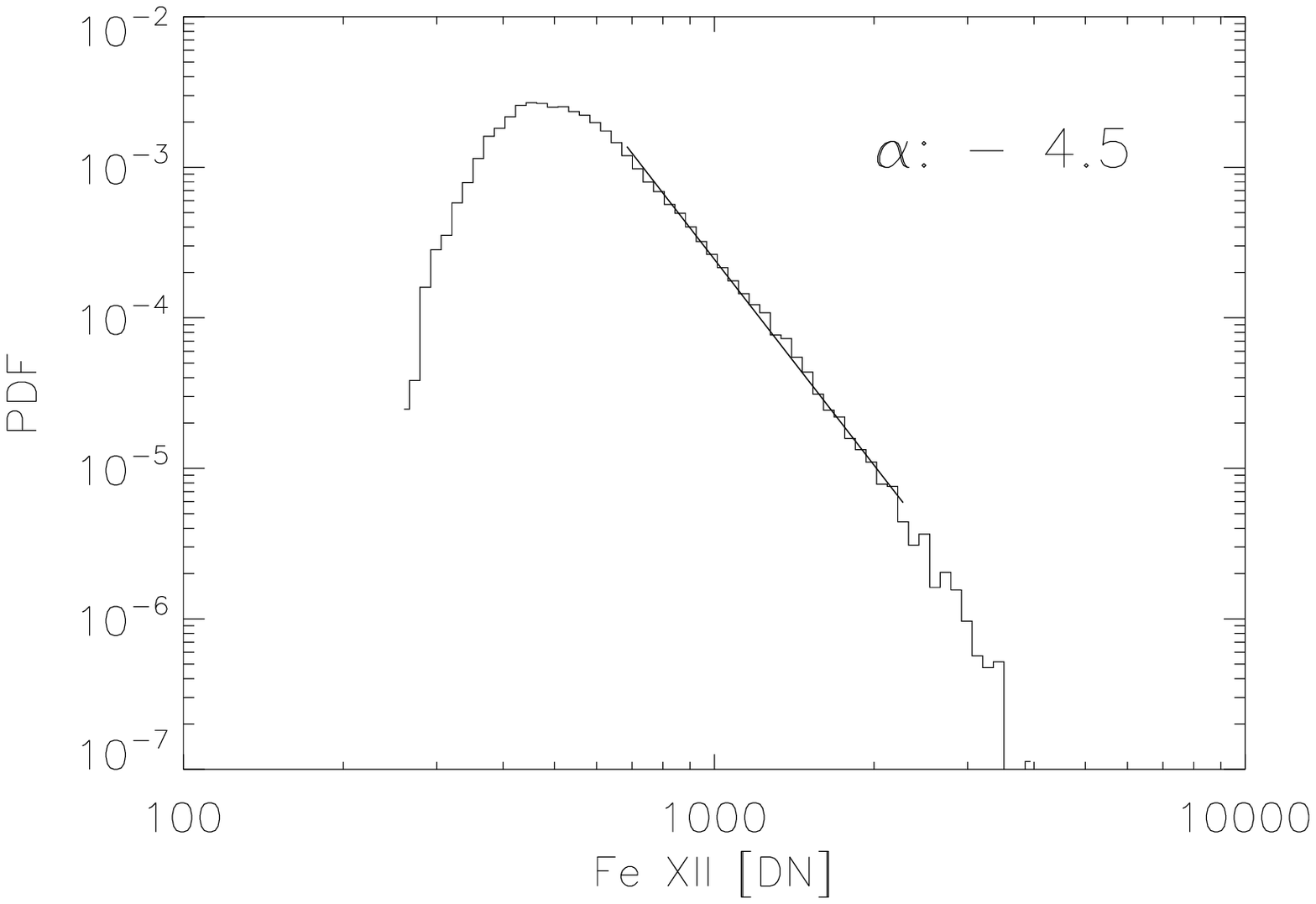}
  \caption{Top: PDFs for the  \ion{Fe}{ix} (on the right) and \ion{Ne}{viii} (on the left) lines. Bottom:  PDFs for the \ion{Fe}{xii} (on the right) and \ion{Mg}{x} (on the left) lines.}
             \label{Fig3}%
   \end{figure*}
For the theoretical calculation of the line intensities, we used the
CHIANTI \citep[v. 5.2,][]{dere97, landi06} atomic database and software,
adopting the \cite{mazzotta98} ion fractions and  photospheric element
abundances \citep{grevesse98}. We simulated the measured data numbers (DN) from the
spectrometers by assuming observations with a 1 arcsec slit and 1s
cadence and using the standard calibration software available in the
SolarSoft IDL package. In the case of EIT, we simulated the total emission
in the channel by integrating the emission from all lines in the
waveband, as predicted by the CHIANTI atomic database, taking into
account the effective area of the instrument. 
The \ion{Fe}{ix} 171 channel of EIT in particular, is affected by significant
contributions from the slightly hotter \ion{Fe}{x} 174.53~\AA~ line, which therefore alters its response to the heating function. The 195 channel has a hot component due to \ion{Fe}{xxiv} 192.03~\AA.

\noindent Figure \ref{GT} shows the contribution functions
$G(T)$ for all lines (top) and the EIT response function for the
channels (bottom) used in this work. The  $G(T)$ functions of the lithium isoelectronic lines in the top plot can be identified by their asymmetry towards high temperatures.
We anticipate in our results that this a high temperature tail will increase the number of
events with low intensity.
% statistics of the low intensities.

\section{Line intensity distributions}
\subsection{``Warm'' line properties}

We investigate the intensity distribution of 
``warm'' coronal lines, i.e., those lines that form at the
average temperature of the corona (1--2~MK).

Figure \ref{Fig3} shows the PDFs for \ion{Fe}{ix} and
\ion{Ne}{viii} 
($\log~T_{\rm max} = 5.8$; top panels), \ion{Fe}{xii}, and
\ion{Mg}{x} ($\log~T_{\rm max} = 6.1$; bottom panels). 
The left panels in the Fig. show the PDFs from the Li-like lines. 
The power-law behaviour of the heating function was transmitted to the
predicted PDFs of each wavelength, although, in each case the index
$\alpha$ was far smaller than the $-1.7$ value of the heating function.
%% Power law behaviour can be identified for each wavelength, similar to 
%% Similar to the heating function, all the PDFs follow a power law,
%% but with an index $\alpha$ much smaller than -1.7. 
This behaviour is consistent with the findings of \cite{parenti06}.
The minimum value of $\alpha$ was reached for the two Li-like lines,
because of the increase in the number of strands emitting a weak intensity; this was caused by the extension of their contribution function to higher temperatures.

To confirm that high temperature tails produced the lower
$\alpha$ values for Li-like ions, the
contribution functions 
were truncated at high temperature and the calculations repeated. The
PDFs were found to then have  $\alpha$ values that were consistent with
the \ion{Fe}{ix} and \ion{Fe}{xii} lines, confirming our hypothesis.
%% to the corresponding distribution for the non-Li isoelectronic line.
%% This behavior of the Li-like line further show how weakly reliable
%% this technique may be in this range of temperatures. 
This comparison demonstrated that the Li-like ions had a
different response to the heating function than other ions formed at
the same temperature, due to the high temperature tails of
the contribution functions.

\subsection{Imager versus spectrometer}

\begin{figure}[h]
  \centering
\includegraphics[scale=.5 ]{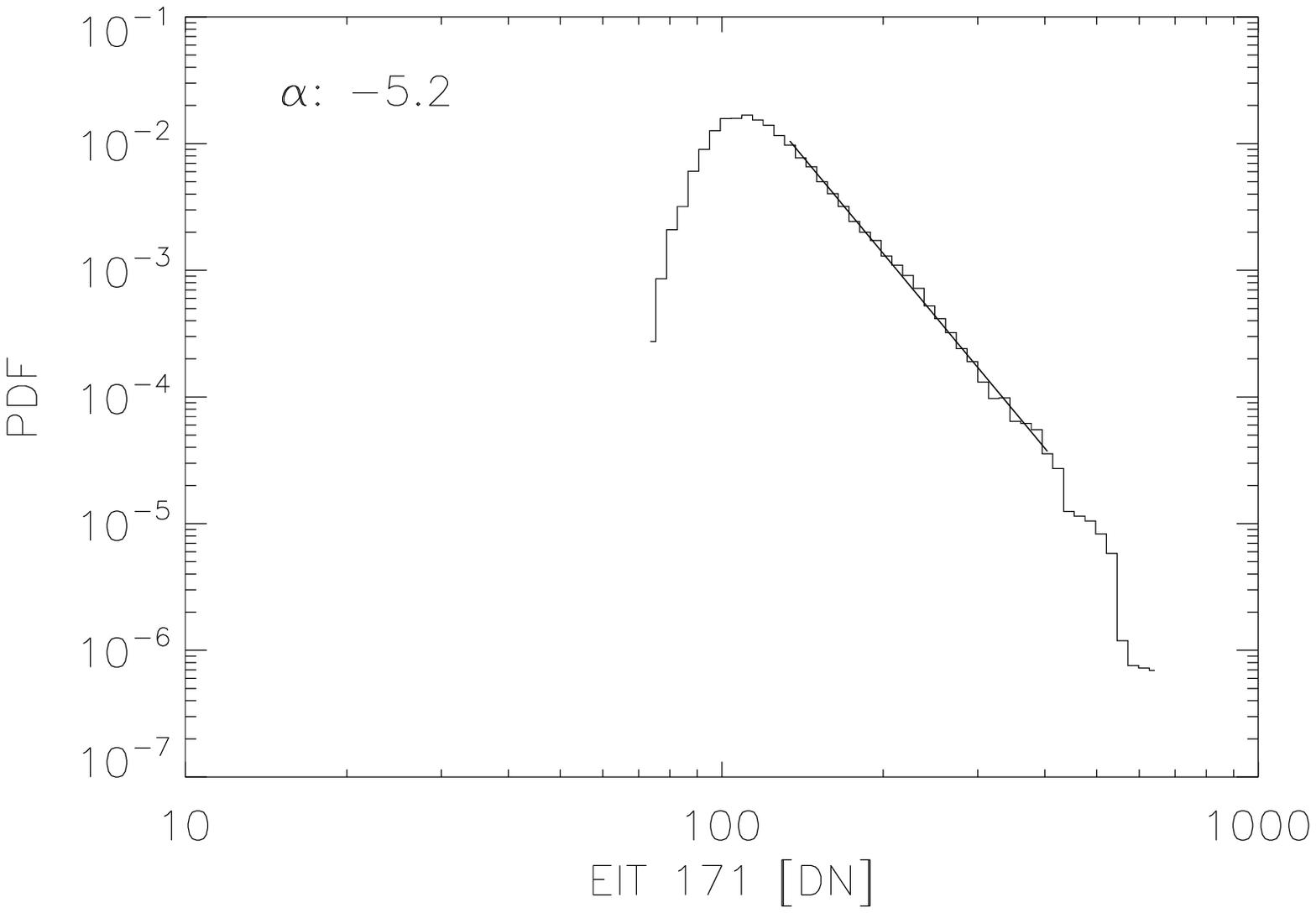}\\
\includegraphics[scale=.5]{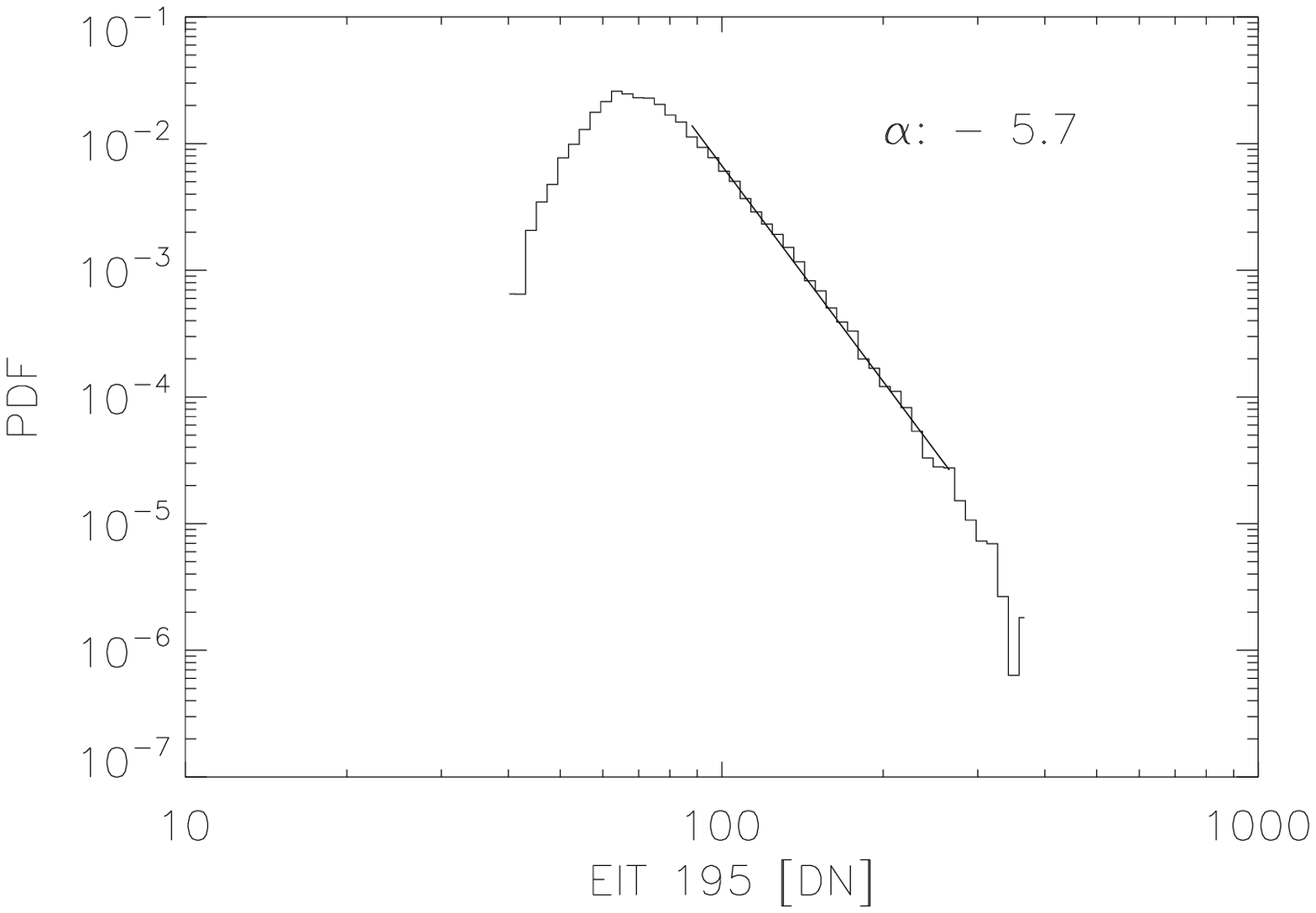}
\caption{PDF for the counts in the EIT 171 (top) and 195 (bottom) channels.}\label{fig4}
\end{figure}

We then investigated weather the difference in the bandwidth of the instrument could have an effect on the  shape of the PDF.  We simulated the EIT intensities in 
the 171 and 195 channels, which were compared with the results presented in
the previous section (Figure \ref{fig4}) for the 171 and 195 emission
lines observed by Hinode/EIS.

Comparing Figs.~\ref{Fig3} and \ref{fig4}, the PDFs for the EIS lines
have indices higher than those derived for the two   EIT channels. We
have therefore demonstrated  that observing emission lines with a wide
band imager can affect the PDF of the emission line.
%Unfortunately, 
We note that the EIS \ion{Fe}{ix} line is situated at the border of the
instrument waveband for which the response is low.
The predicted DN values are therefore very small and the line is not
useful in practice for the study of small-scale brightenings.

%%{\bf The indices for the \ion{Fe}{ix} intensities shown on top of Figures \ref{Fig3} and \ref{fig4} result to be quite similar, even if the simulated data are from narrow  and large band instruments. 

To compare with the spectrometer lines,
the intensities from the EIT channels  were calculated by assuming a 1s
exposure time, which is generally far lower than true EIT exposure times. For a more realistic case, we also calculated  the
PDFs assuming $60~$s exposure times (similar exposure times are
reached during the high cadence EIT ``Shutterless''
program\footnote{See http://sidc.oma.be/EIT/High-cadence.}), and
verified that the PDFs retained the same power-law indices as for the
1s case.

\subsection{``Hot'' line properties}

By hot lines, we indicate those lines that form at temperatures higher
than the average coronal temperature.

\begin{figure}[h]
\includegraphics[scale=0.5]{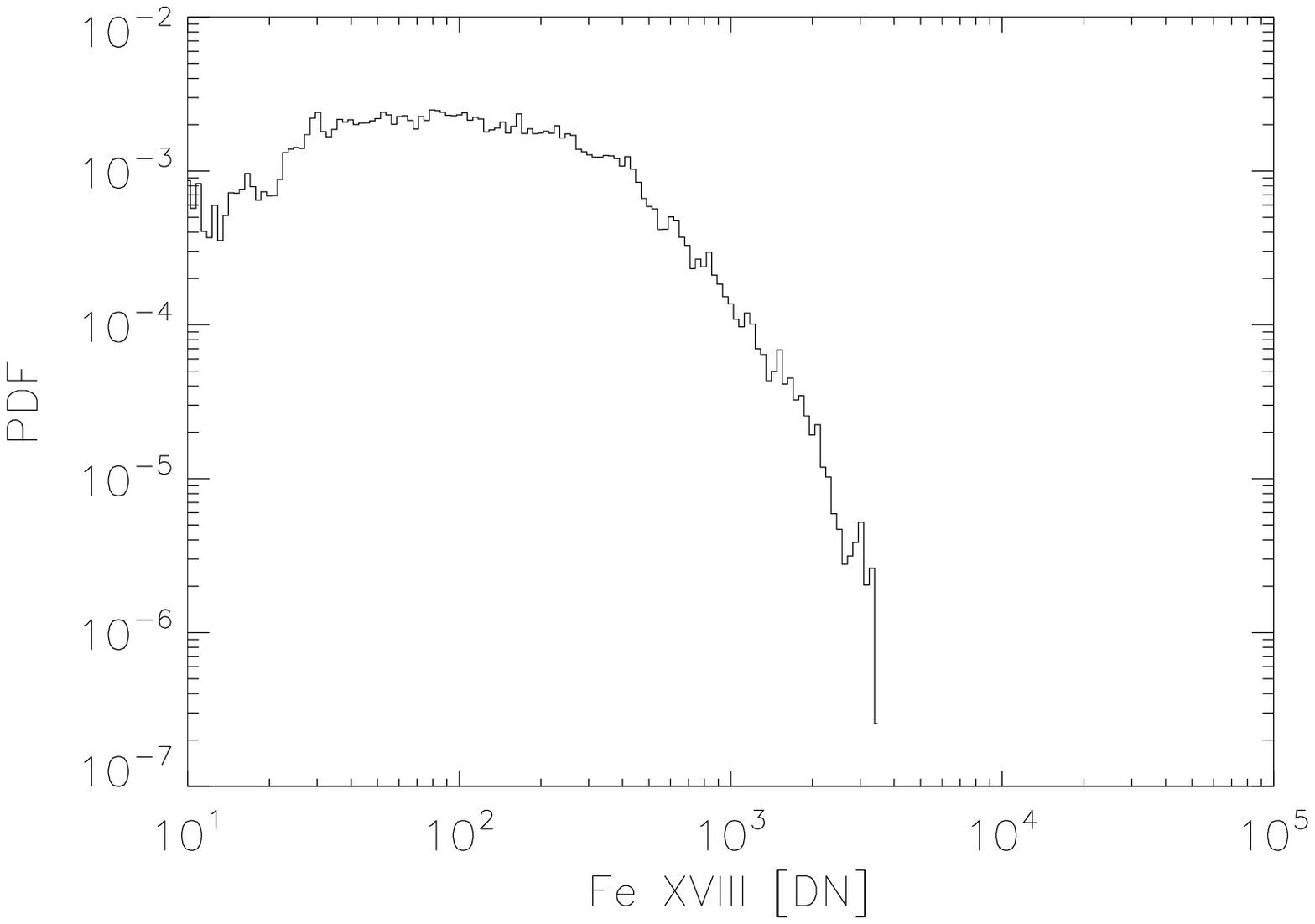}\\
\includegraphics[scale=0.5]{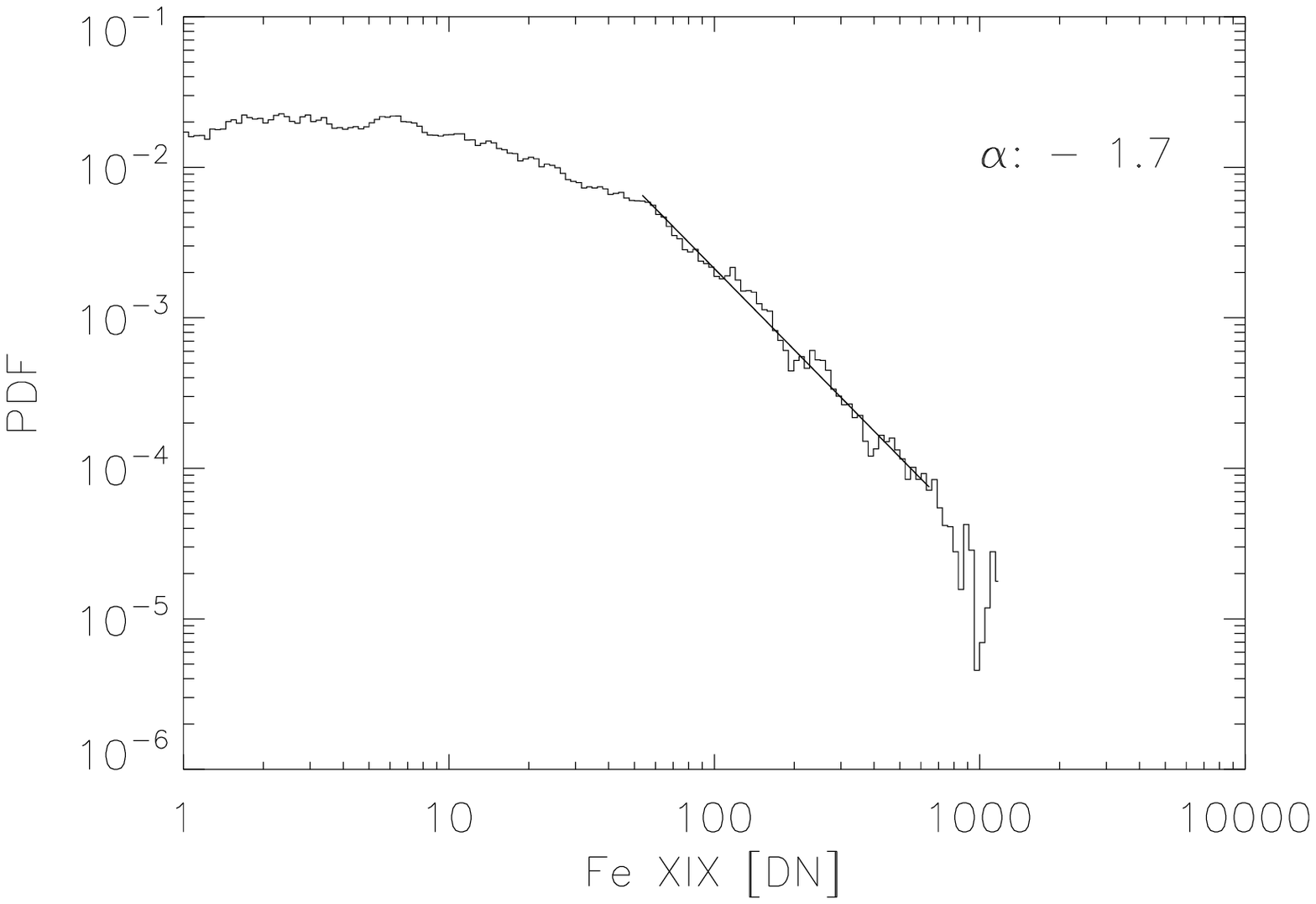}
\caption{PDFs of \ion{Fe}{xviii} (top) and \ion{Fe}{xix} (bottom).}\label{feca}
\end{figure}

The forbidden lines \ion{Fe}{xviii} 974.86 \AA~ and \ion{Fe}{xix}
1118.06 \AA~ are important because they appear at long UV wavelengths
close to cooler lines from the transition region and chromosphere.
A spectrometer observing these lines will therefore observe simultaneously  lines
from distinct regions of the solar atmosphere, a feature valuable for
understanding the multithermal nature of coronal structures.
At $\log~T =6.8$ \ion{Fe}{xviii} forms, close to the DEM peak (see
Figs. \ref{figem} and \ref{GT}),  where the loop threads can cool
by both conduction and  radiation. \ion{Fe}{xix} forms during the cooling phase dominated by thermal conduction only ($\log~T =6.9$). 
Figure \ref{feca} shows the PDFs for these lines (\ion{Fe}{xviii} at the top and \ion{Fe}{xix} at the bottom). The PDF for  \ion{Fe}{xviii} does not have a power-law distribution, while \ion{Fe}{xix} reproduces well  the statistical behaviour of the heating function.
The \ion{Fe}{xix} line was also modelled by \cite{parenti06}, who
found similar results. This finding is then a positive test for our
heating function, which, in contrast to \cite{parenti06}, has been completely synthesised.
Unfortunately there are no strong observed EUV Li-like lines that form
at the same temperatures as \ion{Fe}{xviii} and \ion{Fe}{xix}. For this reason, we cannot directly compare with the results shown in Fig. \ref{feca}.

\begin{figure}[h]
\includegraphics[scale=0.5]{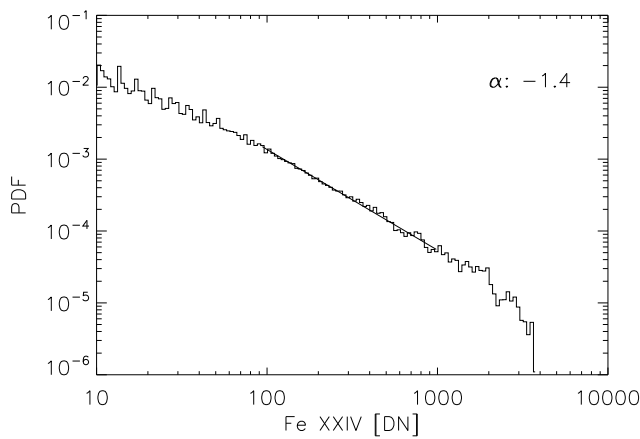}
\caption{The PDF of \ion{Fe}{xxiv}.} \label{fe24}
\end{figure}

As a final test, we compiled the intensity distribution for the most prominent line of the Li-like
\ion{Fe}{xxiv} ion, which exibits a key flare doublet in the EIS
wavelength bands at 192.03~\AA~ and 255.11~\AA\ and whose emission peaks at $\log~T=7.24$ \citep{young07b}. If nanoflares are
responsible for the heating in ARs, we would expect faint, non flaring,
emission at these temperatures \citep[e.g.][]{patsourakos06}. In the
simulation used here, 
the loop system has low DEM values at these high temperatures (Fig.
\ref{figem}), but emission is still found at the \ion{Fe}{xxiv}
line, which will be 
completely formed during the conduction phase.  For the final reason, we would expect that the high temperature wing of the line would not affect the slope of the PDF.
Figure \ref{fe24} shows the resulting PDF. As expected from the previous results, the distribution follows a power-law and the index $\alpha$ is close to that of the heating function
but in fact a little higher, in contrast to results for the other ions that have
been studied. 
We believe that this is because the line forms at the very end of 
the heating function energy range, for which the statistics are probably too small 
to be representative (see also the drop of the  $\mathrm{DEM}(T)$ at high temperatures in Fig. \ref{figem}). This may be a limit to our diagnostic method.

%We note that, unlike for \ion{Ne}{viii} and \ion{Mg}{x}, the high
%temperature tail of the \ion{Fe}{xxiv} is found beyond high
%temperature limit of the DEM, and so it does not skew the derived
%PDF. \ion{Fe}{xxiv} may be thus a valid ion for estimating the index of
%the heating function.

%% At the same time, because the line forms during the conduction phase, we expect  the high temperature wing of the contribution function not to affect the shape of total intensity distribution.

%% Not direct measurements of this line in non flaring ARs exists yet. But with the new Hinode/EIS instrument measuring this line, and the future SDO/AIA  observing \ion{Fe}{xxiii}, the nanoflare hypothesis may be tested. 

\section{Conclusions}

In this work, we have modelled the statistical behaviour of emission line
intensities from a coronal loop, whose heating function is distributed
as a log-normal distribution in energy.
%% In this work we have tested the statistical behavior of a loop
%% emission  as reaction to an heating having the energy distributed
%% selon a log-normal.  
We investigated the plasma response for a wide range  of
temperatures, by simulating  measurements from SOHO/SUMER, Hinode/EIS,
and SOHO/EIT lines, and comparing with the earlier work of
\cite{parenti06}.
% We  compared and extend here out previous results. 
Our results may be summarised as follows: 

\begin{itemize}

\item The synthesised heating function used in this work reproduces the statistical behaviour of a similar function derived by a model  used in a previous work \citep{parenti06}. This would support the use of a modified version of this function in further tests with heating distributions of different power-law indices.

\item We confirm the previous results of \cite{parenti06} that the power-law index of the heating function is preserved only  by the intensity
 distributions of hot lines ($\log\,T\ge6.9$).
%% in finding the power law's index of the intensity distributions similar to that of the heating function only for the high temperature lines.

\item The  shape of the PDF of the line intensities depends not only 
 on the temperature of the line formation  but also on the
 iso-electronic sequence of the emitting ion. In particular, we have
 shown  that the ``warm'' Li-like lines are inappropriate for this
 type of diagnostic, due to the high temperature tail of their
 contribution functions.

\item The behaviour of Li-like \ion{Fe}{xxiv} is not, however, 
 compromised by the high temperature tail of the contribution
 function and the index of the PDF is close
%since the model DEM does not extend to such high temperatures. 
%The \ion{Fe}{xxiv} intensity distribution power-law index is indeed 
%  similar 
to that of the heating function.

\item We identify a weakness in the use of imaging instruments
for statistical studies of coronal heating:
their  wavelength bands generally  contain more than one strong emission line that can affect the PDF.

%% In terms of statistical properties, the behavior of hot Li-like lines seems to be consistent with those of other iso-electronic sequences. However, some discrepancies exist and may require further study.

\end{itemize}
{

In light of the present results, the high temperature channels on AIA (\ion{Fe} {xxiii} 133 \AA) and Solar Orbiter (EUI) or the flaring spectroscopic lines on Solar Orbiter/EUS could be an important source of information for the coronal heating problem. At the same time, sensible contributions from other lines in the large band instruments need to be carefully investigated.

The results presented here were obtained using a  simple, hydrodynamical
model, halthough it has been shown that the model could reproduce   the general properties of the plasma well.
In the future, we will search for further confirmation of our findings by
using an evolved version of our model.

\begin{acknowledgements}

SP would like to thank David Berghmans for the fruitful discussion.
SP acknowledge the support from the Belgian Federal Science Policy Office through the ESA-PRODEX programme.
This work was partially supported by the International Space Science Institute in the framework of an international working team (n. 108).
CHIANTI is a collaborative project involving the NRL (USA), RAL (UK), MSSL (UK), the Universities of Florence (Italy) and Cambridge (UK), and George Mason University (USA).
\end{acknowledgements}

%\begin{thebibliography}{}

%\end{thebibliography}
\bibliographystyle{aa} 
\bibliography{bib}

\begin{thebibliography}{20}
\expandafter\ifx\csname natexlab\endcsname\relax\def\natexlab#1{#1}\fi

\bibitem[{{Aletti} {et~al.}(2000){Aletti}, {Velli}, {Bocchialini}, {Einaudi},
  {Georgoulis}, \& {Vial}}]{aletti00}
{Aletti}, V., {Velli}, M., {Bocchialini}, K., {et~al.} 2000, \apj, 544, 550

\bibitem[{{Aschwanden}(2005)}]{aschwanden05}
{Aschwanden}, M.~J. 2005, {Physics of the Solar Corona. An Introduction with
  Problems and Solutions (2nd edition)} (Pour la Science)

\bibitem[{{Aschwanden} \& {Parnell}(2002)}]{aschwanden02}
{Aschwanden}, M.~J. \& {Parnell}, C.~E. 2002, \apj, 572, 1048

\bibitem[{{Berghmans} {et~al.}(1998){Berghmans}, {Clette}, \&
  {Moses}}]{berghmans98}
{Berghmans}, D., {Clette}, F., \& {Moses}, D. 1998, \aap, 336, 1039

\bibitem[{{Buchlin} {et~al.}(2003){Buchlin}, {Aletti}, {Galtier}, {Velli},
  {Einaudi}, \& {Vial}}]{buchlin03}
{Buchlin}, E., {Aletti}, V., {Galtier}, S., {et~al.} 2003, \aap, 406, 1061

\bibitem[{{Cargill}(1994)}]{cargill94}
{Cargill}, P.~J. 1994, \apj, 422, 381

\bibitem[{{Christe} {et~al.}(2008){Christe}, {Hannah}, {Krucker}, {McTiernan},
  \& {Lin}}]{christe08}
{Christe}, S., {Hannah}, I.~G., {Krucker}, S., {McTiernan}, J., \& {Lin}, R.~P.
  2008, \apj, 677, 1385

\bibitem[{{Culhane} {et~al.}(2007){Culhane}, {Harra}, {James}, {Al-Janabi},
  {Bradley}, {Chaudry}, {Rees}, {Tandy}, {Thomas}, {Whillock}, {Winter},
  {Doschek}, {Korendyke}, {Brown}, {Myers}, {Mariska}, {Seely}, {Lang}, {Kent},
  {Shaughnessy}, {Young}, {Simnett}, {Castelli}, {Mahmoud}, {Mapson-Menard},
  {Probyn}, {Thomas}, {Davila}, {Dere}, {Windt}, {Shea}, {Hagood}, {Moye},
  {Hara}, {Watanabe}, {Matsuzaki}, {Kosugi}, {Hansteen}, \&
  {Wikstol}}]{culhane07}
{Culhane}, J.~L., {Harra}, L.~K., {James}, A.~M., {et~al.} 2007, \solphys, 243,
  19

\bibitem[{{Delaboudiniere} {et~al.}(1995){Delaboudiniere}, {Artzner},
  {Brunaud}, {Gabriel}, {Hochedez}, {Millier}, {Song}, {Au}, {Dere}, {Howard},
  {Kreplin}, {Michels}, {Moses}, {Defise}, {Jamar}, {Rochus}, {Chauvineau},
  {Marioge}, {Catura}, {Lemen}, {Shing}, {Stern}, {Gurman}, {Neupert},
  {Maucherat}, {Clette}, {Cugnon}, \& {van Dessel}}]{delaboudiniere95}
{Delaboudiniere}, J.-P., {Artzner}, G.~E., {Brunaud}, J., {et~al.} 1995,
  \solphys, 162, 291

\bibitem[{{Dere} {et~al.}(1997){Dere}, {Landi}, {Mason}, {Monsignori Fossi}, \&
  {Young}}]{dere97}
{Dere}, K.~P., {Landi}, E., {Mason}, H.~E., {Monsignori Fossi}, B.~C., \&
  {Young}, P.~R. 1997, \aaps, 125, 149

\bibitem[{{Golub}(2006)}]{golub06}
{Golub}, L. 2006, Space Science Reviews, 124, 23

\bibitem[{{Grevesse} \& {Sauval}(1998)}]{grevesse98}
{Grevesse}, N. \& {Sauval}, A.~J. 1998, Space Science Reviews, 85, 161

\bibitem[{{Hochedez} \& al.(2007)}]{hochedez07}
{Hochedez}, J.~F. \& al. 2007, in ESA SP-641: Second Solar Orbiter Workshop,
  published on CDROM, ""

\bibitem[{{Landi} {et~al.}(2006){Landi}, {Del Zanna}, {Young}, {Dere}, {Mason},
  \& {Landini}}]{landi06}
{Landi}, E., {Del Zanna}, G., {Young}, P.~R., {et~al.} 2006, \apjs, 162, 261

\bibitem[{{Mazzotta} {et~al.}(1998){Mazzotta}, {Mazzitelli}, {Colafrancesco},
  \& {Vittorio}}]{mazzotta98}
{Mazzotta}, P., {Mazzitelli}, G., {Colafrancesco}, S., \& {Vittorio}, N. 1998,
  \aaps, 133, 403

\bibitem[{{Parenti} {et~al.}(2006){Parenti}, {Buchlin}, {Cargill}, {Galtier},
  \& {Vial}}]{parenti06}
{Parenti}, S., {Buchlin}, E., {Cargill}, P.~J., {Galtier}, S., \& {Vial}, J.-C.
  2006, \apj, 651, 1219

\bibitem[{{Patsourakos} \& {Klimchuk}(2006)}]{patsourakos06}
{Patsourakos}, S. \& {Klimchuk}, J.~A. 2006, \apj, 647, 1452

\bibitem[{{Wilhelm} {et~al.}(1995){Wilhelm}, {Curdt}, {Marsch}, {Schuhle},
  {Lemaire}, {Gabriel}, {Vial}, {Grewing}, {Huber}, {Jordan}, {Poland},
  {Thomas}, {Kuhne}, {Timothy}, {Hassler}, \& {Siegmund}}]{wilhelm95}
{Wilhelm}, K., {Curdt}, W., {Marsch}, E., {et~al.} 1995, \solphys, 162, 189

\bibitem[{{Young} \& al.(2007a)}]{young07a}
{Young}, P.~R. \& al. 2007a, in ESA SP-641: Second Solar Orbiter Workshop,
  published on CDROM, ""

\bibitem[{{Young} {et~al.}(2007b){Young}, {Del Zanna}, {Mason}, {Dere}, {Li},
  {Lini}, {Doschek}, {Brown}, {Culhane}, {Harra}, {Watanabe}, \&
  {Hara}}]{young07b}
{Young}, P.~R., {Del Zanna}, G., {Mason}, H.~E., {et~al.} 2007b, \pasj, 59, 857

\end{thebibliography}

\end{document}